\documentclass[%
reprint,
 amsmath,amssymb,superscriptaddress,
  aps,
longbibliography
]{revtex4-1}
\usepackage{xparse}
\usepackage{appendix}
\usepackage{graphicx}
\usepackage{dcolumn}
\usepackage{bm}
\usepackage{hyperref}
 \UseRawInputEncoding
\renewcommand{\d}[2]{\ensuremath{\frac{\text{d} #1}{\text{d} #2}}}

\newcommand{\pd}[2]{\ensuremath{\frac{\partial #1}{\partial #2}}}

\usepackage{slashed}
\renewcommand{\vec}[1]{\mathbf{#1}}

\usepackage{bm}
\newcommand{\vect}[1]{\boldsymbol{\mathbf{#1}}}

\usepackage{bbold}

\usepackage{xcolor}

\begin{document}
\title{Non-Abelian bosonization in a $(3+1)$-d Kondo semimetal via quantum anomalies}

	\author{Colin Rylands}
\affiliation{Joint Quantum Institute and
 Condensed Matter Theory Center, University of Maryland, College Park, MD 20742, USA}
\author{Alireza Parhizkar}
\affiliation{Joint Quantum Institute and
	Condensed Matter Theory Center, University of Maryland, College Park, MD 20742, USA}
\author{Victor Galitski}
\affiliation{Joint Quantum Institute and
	Condensed Matter Theory Center, University of Maryland, College Park, MD 20742, USA}

\date{
    \today
 }

\begin{abstract}
Kondo lattice models have established themselves as an ideal platform for studying the interplay between topology and strong correlations such as in topological Kondo insulators or Weyl-Kondo semimetals. The nature of these systems requires the use of non-perturbative techniques which are few in number, especially in high dimensions.  Motivated by this we study a model of Dirac fermions  in $3+1$ dimensions coupled to an arbitrary array of spins via a generalization of functional non-Abelian bosonization. We show that there exists an exact transformation of the fermions which allows us to write the system as decoupled free fermions and interacting spins. This decoupling transformation consists of a local chiral, Weyl and Lorentz transformation parameterized by solutions to a set of nonlinear differential equations which order by order takes the form of Maxwell's equations with the spins acting as sources. Owing to its chiral and Weyl components this transformation is anomalous and generates a contribution to the action. From this we obtain  the effective action for the spins and expressions for the anomalous transport in the system. In the former we find that the coupling to the fermions generates kinetic terms for the spins, a long ranged interaction and a Wess-Zumino like term. In the latter we find  generalizations of the chiral magnetic and Hall effects.  
\end{abstract}
\date{\today}

\maketitle


\section{Introduction}Quantum impurity models are a prime example of strongly correlated condensed matter systems, facilitating our understanding of many physical phenomena including the ubiquitous Kondo effect. When many impurities are present as is the case in a Kondo lattice,  hybridization between the localized spins and the itinerant fermions leads to  a variety of effects including  rich Heavy fermion physics~\cite{hewson_1993, coleman_2015}. More recently, the Kondo lattice has been the focus of attention for its possible topological properties and in particular the potential to study the interplay between topology and strong correlations. In those cases, the strong correlations result in the emergence of topological phases of matter including  topological Kondo insulators~\cite{DzeroSunGalitskiColeman, DzeroARCMP} and  Weyl-Kondo semimetals~\cite{Dzsaber, LaiGrefePaschenSi, dzsaber2021giant, ChangColeman}. In the latter case, due to the Kondo effect, the low energy excitations of the system are Weyl fermions. Weyl semimetals are of great interest in and of themselves providing concrete realizations in a solid-state system of physical phenomena  historically associated with particle physics. Chief amongst these is the chiral anomaly, the breaking of classical chiral symmetry~\cite{Adler, BellJackiw, Fujikawa2004Book} in a quantum theory  which gives rise to several distinct transport features in free~\cite{NielsenNinomiya,  WanTurnerVishwanathSavrasov, BurkovBalents, YangLuRab, XuWengWangDaiFang, HalaszBalents, Aji,WengFangZhongBernevigDai, Lv1, Lv2, Xu, Huang,SonSpivak, GoswamiTewari} and interacting systems~\cite{RainesGalitski, RylandsParhizkar}. 

Motivated by these considerations and in particular the effects of quantum anomalies in strongly correlated systems, we study a system of $3+1$-dimensional Dirac fermions coupled to an arbitrary array of spins.  In lower dimensions there exist many analytic, non-perturbative or exact  techniques to study  Kondo models including conformal field theory~\cite{Affleck,FRADKIN2}, Bethe Ansatz~\cite{NatanRMP, TWAKM, CR1,CR2,CR3,PRA1, PRAA} and bosonization~\cite{GiamarchiBook, GogolinNerseyanTsvelikBook,FRADKIN1,Tsvelik, GoswamiSi,PhysRevX.5.021017}. In higher dimensions there is a lack of non-perturbative techniques and typically  a slave particle approach is adopted~\cite{Read_1983, colemanPRB, hewson_1993, coleman_2015, ChangColeman}. In this work we will take an alternative approach. Our method will follow that of the anomaly based formulation of functional bosonization, appropriately generalized to the present situation. In the original formulation one considers  Dirac fermions in $1+1$ dimensions, with either Abelian or non-Abelian symmetry,  coupled to some fluctuating field e.g. a Hubbard Stratonovich field. The fermions are decoupled from this field via a judiciously chosen local chiral and gauge rotation after which the system consists of free fermions and a decoupled fluctuating field whose effective action is calculated using the chiral anomaly \cite{Gamboa, Furuya, Naon, LeeChen}. The remaining fermonic degrees of freedom are easily integrated out resulting in an effective bosonic theory.  We shall follow the same procedure for our system, with the spin-momentum locking of Dirac fermions necessitating a non-Abelian formulation. In addition, the increase in dimensions shall make the formalism more complex and ultimately will not end in an exact solution which can be the case in lower dimensions~\cite{GiamarchiBook, GogolinNerseyanTsvelikBook}. In spite of this, the approach provides us access to some exact results including that of the anomalous transport in the system.  

{{This paper is organized as follows: In Section II we introduce the model and discuss the relation between our method and the standard chiral anomaly treatment of Weyl semimetals. In section III we formulate the decoupling transformation and  show how it can reduce the system to free fermions and a decoupled interacting spin system. In the subsequent section we present an iterative scheme for finding this transformation.  In section V we calculate the contributions of the chiral and Weyl anomalies to the action for our model. These are then used in section VI to determine the low energy effective action for the spin system. Following this we determine the anomalous transport, finding a modification to the quantum Hall current and a chiral magnetic effect (CME) due to the fluctuations of the spins. In the last section we summarize and conclude. }}



\section{Model} Our system is described by the path integral
\begin{eqnarray}\label{Z}
Z=\int \mathcal{D}\psi\mathcal{D} \bar{\psi}\mathcal{D} \vec{S} \, e^{i\mathcal{S}[\psi,\bar{\psi},\vec{S}]}
\end{eqnarray}
where the action is $\mathcal{S}=\mathcal{S}_D+\mathcal{S}_\text{spin}$ with
\begin{eqnarray}\label{SD}
\mathcal{S}_D\!=\!\!\int d^4 x \,\bar\psi(x)i\gamma^\mu\left[\partial_\mu-ie A_\mu(x)-i J \gamma_5 S_\mu(x)\right]\psi(x).
\end{eqnarray}
Here  $\bar{\psi}(x),\psi(x)$ are four component Dirac fermions, (spinor indices suppressed for the moment) describing the low energy sector of a semimetal. They are coupled via a spin exchange of strength $J$ to a system of spins, $S_\mu(x)=(0,\vec{S}(x))$, governed by the action $\mathcal{S}_{\text{spin}}$ and a gauge field $A_\mu(x)$.  Specific cases of this model have been studied previously including the Kondo effect  for a single impurity~\cite{MitchellFritz} and also the Ruderman-Kittel-Kasuya-Yosida  (RKKY) interaction for two impurities~\cite{Chang}. Our approach here does not depend upon the form of $\mathcal{S}_{\text{spin}}$, $\vec{S}(x)$ can be an arbitrary vector field, classical or quantum as in~\eqref{Z}.  

{{This action, \eqref{SD}, can also be viewed as that of a low energy description of a semimetal subject to strain fields \cite{Vozmediano1, Vozmediano2}. The strain induces chiral gauge fields in the action for which the field $\vec{S}(x)$ plays the role of a vector potential e.g. a rotational strain will induce a chiral magnetic field with strength $\vec{\nabla}\times \vec{S}(x)$. Our results are also applicable to these strained semimetals however in the remainder of the paper we restrict the terminology and perspective to that of the Kondo semimetal.}}

We shall perform a rotation on the fermionic degrees of freedom such that $\vec{S}(x)$ is removed from $\mathcal{S}_D$.  For a simple Weyl semimetal, when $\vec{S}$ is constant this  is easily carried out via a local chiral gauge transformation $$\psi(x)\to e^{i\gamma_5 \vec{x}\cdot \vec{S}}\psi(x),~\bar{\psi}(x)\to \bar{\psi}(x) e^{i\gamma_5 \vec{x}\cdot \vec{S}}$$ which transforms the action $\mathcal{S}_D\to \mathcal{S}_D'=\int dx \,\bar{\psi}(x)\,i\gamma^\mu\left[\partial_\mu-ie A_\mu(x)\right]\psi(x)$~\cite{ZyuzinBurkov}. Such a transformation is known to be anomalous \cite{Adler, BellJackiw}, meaning that it results in a nontrivial Jacobian in the path integral measure i.e. $$\mathcal{D}\psi\mathcal{D} \bar{\psi}\to \mathcal{D}\psi \mathcal{D} \bar{\psi}\, e ^{i \mathcal{S}_\mathcal{A}}.$$ The anomalous contribution to the transformed action, $\mathcal{S}_\mathcal{A}$ is straightforwardly calculated using the method of Fujikawa~\cite{Fujikawa, FujikawaErrata}. It takes the form of an axion-like term
\begin{eqnarray}\label{Weyl}
\mathcal{S}_\mathcal{A}=J\frac{e^2 }{4\pi^2}\int d^4 x \,\epsilon^{\nu\mu\rho\sigma} S_\mu A_\nu \partial_\rho A_\sigma. 
\end{eqnarray}
Using this one may then calculate the anomalous transport of the fermions by varying the action with respect to $A_\mu(x)$. This gives the Hall current $\left<\vec{j}\right>=J\frac{e^2}{2\pi}\vec{S}\times \vec{E}$ and density $\left<\rho\right>=J\frac{e^2}{2\pi}\vec{S}\cdot \vec{B}$ where $\vec{E}, \,\vec{B}$ are external  electric and magnetic fields. When $S_0\neq0$  inversion symmetry is broken and  the chiral magnetic effect occurs in the presence of a magnetic field ~\cite{FukushimaKharzeevWarringa, ChenWuBurkov, ZyuzinBurkov}. 

When $\vec{S}$ is not constant a local chiral gauge transformation is no longer sufficient to decouple the fermions from the spins. As we will show below it is still possible but the transformation which does this is non-Abelian, consisting of a combination of local chiral, Weyl and Lorentz transformations. The first two of these are anomalous and will generate a contribution to action, which includes interactions between the spins and also provide a route to calculating the exact anomalous transport. For simplicity we restrict ourselves to the zero chemical potential and zero chiral chemical potential 
{(i.e. $S_0=0$ which is also the case for a stained semimetal) }
however both can be straightforwardly accommodated within our approach.  {{In addition we treat only the zero temperature and infinite volume case.}}


\section{Decoupling Transformation} For arbitrary $\vec{S}(x)$ the appropriate decoupling transformation is $\psi(x)\to U(x)\psi(x)$ and  $\bar{\psi}(x)\to \bar{\psi}(x)\overline{U}(x)$ with 
\begin{eqnarray}\label{M}
U(x)=e^{i\gamma_5\phi(x)+\Omega(x)+i\gamma_5\mathcal{F}_{\mu\nu}(x)\sigma^{\mu\nu}}
\end{eqnarray} 
and $\overline{U}(x)=\gamma_0U^\dag(x)\gamma_0$ where $\sigma^{\mu\nu}=[\gamma^\mu,\gamma^\nu]/2$ are the generators of Lorentz transformations in the spinor representation. {{Heuristically, we can understand the form of this transformation in the following way. We envision an array of spins, each of arbitrary length and orientation.  Using a local Lorentz transformation we can locally rotate to a frame where the spins are parallel but of differing length. They can then be rescaled in length to be the same using the local Weyl transformation and following this they can then be decoupled through the local chiral transformation. }}

More specifically, the real functions $\phi(x),~ \Omega(x)$, $\mathcal{F}_{\mu\nu}(x)$  which  parameterize the local chiral, Weyl and Lorentz transformations respectively  are determined by solving
\begin{eqnarray}\label{Condition}
i\left[\slashed{\partial} U(x)\right] U^{-1}(x)=J\gamma_5\slashed{S}(x),
\end{eqnarray}
where we have employed Dirac slash notation; $\slashed{C}\equiv\gamma^\mu C_\mu$. As opposed to the case of constant $\vec{S}$, the non-Abelian nature of $U(x)$ now makes this a non trivial task  which we shall address in the next section.  
Using this transformation in \eqref{SD} the action is transformed as $\mathcal{S}_D\to \mathcal{S}'_D$, 
\begin{eqnarray}\label{Sprime}
\mathcal{S}'_D&=&\int d ^4 x\, \bar{\psi}(x)\overline{U}(x)\gamma^\mu U(x) \left[\partial_\mu-iA_\mu\right]\psi(x)\\
&=&\int d ^4 x\, \bar{\psi}(x)e^{2\Omega(x)}\Lambda^\mu_\nu(x) \gamma^\nu \left[\partial_\mu-iA_\mu\right]\psi(x). 
\end{eqnarray}
In the second line we have introduced $\Lambda^\mu_\nu(x) =[e^{i\gamma_5\mathcal{F}_{\alpha\beta}(x)\omega^{\alpha\beta}}]^\mu_{\,\nu}$ with $\omega^{\alpha\beta}$ being the generators of Lorentz transformations in the vector representation. We then perform a coordinate  transformation $x\to y(x)$ such that,
\begin{eqnarray}\label{coord}
\d{y^\mu(x)}{x^\nu}=e^{-2\Omega(x)/3}\Lambda^\mu_\nu(x)
\end{eqnarray}
This transformation does not have unit determinant due to the $\Omega(x)$ term, however the coefficient in the exponent is chosen such that the Jacobian of this transformation is cancelled. Ultimately, we obtain
\begin{eqnarray}
\mathcal{S}'_D=\int d^4 y\,\bar{\psi}(y)\gamma^\mu \left[\partial_\mu-i\tilde{A}_\mu\right]\psi(y). 
\end{eqnarray}
Which is the action of free Dirac fermions coupled to a rotated and rescaled gauge field 
\begin{eqnarray}\label{tildeA}
 \tilde{A}_\mu(y)=e^{2\Omega(x)/3}\Lambda^\nu_\mu(x) A_\nu(x).
\end{eqnarray}  

In the absence of the gauge field the fermion and spin system have been decoupled. Therefore, provided that a solution to  \eqref{Condition} exists our path integral is transformed to
\begin{eqnarray}
Z=\int \mathcal{D}\psi\mathcal{D} \bar{\psi}\mathcal{D} \vec{S} \, e^{i\mathcal{S_D'}[\psi,\bar{\psi}]+i\mathcal{S}_\mathcal{A}[\vec{S}]+i\mathcal{S}_\text{spin}[\vec{S}]}
\end{eqnarray}
where $\mathcal{S}_\mathcal{A}$ comes from the Jacobian of the chiral and Weyl transformations which depends upon $\vec{S}(x)$ and $A_\mu(x)$. 

{{We note that the defining equation for the transformation \eqref{Condition} is a Dirac equation of the type which the untransformed  fermion obeys. In the noninteracting  case, $J=0$, $U(x)$ should reduce to the identity and so we can view it as the operator which locally transforms the field from the Heisenberg to the interaction picture. We expect on general grounds that this is generically possible to implement. In contrast to the standard procedure however we carry out this transformation in the path integral which turns out to be anomalous. }}



\section{Iterative Solution}
The task now is to solve \eqref{Condition} for $\phi(x),\Omega(x)$ and $\mathcal{F}_{\mu\nu}(x)$ in terms of $\vec{S}$. To do this we introduce  $\mathcal{E}_i=\mathcal{F}_{0 i}$ and $\mathcal{B}_i=-\frac{1}{2}\epsilon_{ijk}\mathcal{F}^{jk}$ with Latin indices reserved for spatial components. Inserting this form into \eqref{Condition} and using standard vector calculus identities we obtain a set of non linear differential equations for our unknown functions, $\phi,\Omega,\vect{\mathcal{E}}$ and $\vect{\mathcal{B}}$ which resemble the equations for a driven two level system~\cite{Galitski, Gangopadhyay_2010}. We solve these by expanding in powers of $J$ i.e.  $\phi(x)=\sum_{n=1}^\infty J^n\phi^{(n)}$ and proceeding iteratively. 

The leading order equations resemble Maxwell's equations with magnetic source terms. Therein, $\vect{\mathcal{E}}^{(1)}$ and $\vect{\mathcal{B}}^{(1)}$ play the role of pseudo-electric and pseudo-magnetic fields and  $\vec{S},~\phi^{(1)}, ~\Omega^{(1)}$ provide the sources,
\begin{eqnarray}\label{Maxwells1}
\partial_t \vect{\mathcal{E}}^{(1)}-\vect{\nabla}\times\vect{\mathcal{B}}^{(1)}&=&\vec{S}-\vect{\nabla}\phi^{(1)}\\
\partial_t \vect{\mathcal{B}}^{(1)}+\vect{\nabla}\times\vect{\mathcal{E}}^{(1)}&=&-\vect{\nabla}\Omega^{(1)}\\\label{Maxwells2}
\vect{\nabla}\cdot \vect{\mathcal{B}}^{(1)}=\partial_t\Omega^{(1)},~\vect{\nabla}\cdot\vect{\mathcal{E}}^{(1)}&=&\partial_t\phi^{(1)}.
\end{eqnarray}
The solution of these equations is known from classical electromagnetism; $\phi^{(1)}(x)=\vect{\nabla}\left[G*\vec{S}(x)\right]$, ~$\vect{\mathcal{E}}^{(1)}(x)=\partial_t\left[G*\vec{S}(x)\right]$ and $\vect{\mathcal{B}}^{(1)}(x)=-\vect{\nabla}\times \left[G*\vec{S}(x)\right]$ in addition to $\Omega^{(1)}(x)=0$. Here $G(x)$  is the Green's function for the d'Alembertian, $[\partial_t^2-\vect{\nabla}^2]G(x)=\delta^{(4)}(x)$ and $*$ stands for convolution, $G*\vec{S}(x)=\int d^4z G(x-z) \vec{S}(z)$. Note that since  $\Omega^{(1)}(x)$ vanishes, no Weyl transformation is required at this order and \eqref{Maxwells1}-\eqref{Maxwells2} reduce to Maxwell's equations without magnetic monopole terms. 

We may express this linearized solution in a more elegant form. To do this we recall that $G(x)$ can be related to the  Green's function, $\mathcal{G}(x)$, for the massless Dirac equation through $\mathcal{G}(x)\equiv\slashed{\partial} G(x)$. Using this  we have that to linear order
\begin{eqnarray}\label{M1}
U(x)=e^{i J \gamma_5\mathcal{G}*\slashed{S}(x)}.
\end{eqnarray}
The higher order corrections to this, $\vect{\mathcal{E}}^{(n)}$ and $\vect{\mathcal{B}}^{(n)}$ are also solutions to Maxwell's equations but with sources which are determined by the lower order terms.  For example at second order
\begin{eqnarray}\label{Maxwells3}
\partial_t \vect{\mathcal{E}}^{(2)}-\vect{\nabla}\times\vect{\mathcal{B}}^{(2)}&=&\text{Re}[\vec{S}^{(1)}]-\vect{\nabla}\phi^{(2)},\\\label{Maxwells4}
\partial_t \vect{\mathcal{B}}^{(2)}+\vect{\nabla}\times\vect{\mathcal{E}}^{(2)}&=&\text{Im}[\vec{S}^{(1)}]-\vect{\nabla}\Omega^{(2)},\\
\vect{\nabla}\cdot \vect{\mathcal{B}}^{(2)}&=&\partial_t\Omega^{(2)}-\text{Im}[S^{(1)}_{0}],\\
\vect{\nabla}\cdot\vect{\mathcal{E}}^{(2)}&=&\partial_t\phi^{(2)}-\text{Re}[S^{(1)}_{0}]
\end{eqnarray}
where we have introduced $\vect{S}^{(1)}=\vect{X}^{(1)}\times \left[\partial_t+i \vect{\nabla}\times \right]\vect{X}^{(1)}$ and $S_0^{(1)}=\vect{X}^{(1)}\cdot \vect{\nabla}\times \vect{X}^{(1)}$ with $\vect{X}^{(1)}=\vect{\mathcal{E}}^{(1)}+i\vect{\mathcal{B}}^{(1)}$. The solution to these can be found from a straightforward generalization of the linear order solution, i.e. derivative operators acting on terms like $G*S_\mu^{(1)}$. Combining this with \eqref{M1} we have that up to second order  $U(x)= e^{ J \mathcal{G}*\left(J\text{Im}[\slashed{S}^{(1)}]-i\gamma_5(\slashed{S}+J\text{Re}[\slashed{S}^{(1)}]\right)}$. All higher orders proceed along similar lines and we can write the full solution as 
\begin{eqnarray}
U(x)=e^{ J\mathcal{G}*( \text{Im}[\slashed{\mathbb{S}}(x)]-i\gamma_5\text{Re}[\slashed{\mathbb{S}}(x)])}
\end{eqnarray}
where $\mathbb{S}_\mu=\sum_{n=0}^\infty J^n S_\mu^{(n)}$ and $S_\mu^{(0)}= S_\mu$. Matching this to \eqref{M} then gives $\phi(x)=\frac{J}{4}\text{tr}(\mathcal{G}*\text{Re}[\slashed{\mathbb{S}}])$, $\Omega(x)=\frac{J}{4}\text{tr}( \mathcal{G}*\text{Im}[\slashed{\mathbb{S}}])$ and {{ $\mathcal{F}^{\mu\nu}(x)=-\frac{J}{8}\text{tr}[\sigma^{\mu\nu}\mathcal{G}*(\gamma_5\text{Re}[\slashed{\mathbb{S}}]-i\text{Im}[\slashed{\mathbb{S}}])]$.   }}

The corrections to $\mathbb{S}_\mu$ naturally become  more complicated at higher orders. Notably,  they contain an increasing number of derivatives each time,  i.e. $S^{(n)}$ contains at least  $n$  derivatives acting on $\vec{S}$. Accordingly, if for some $n$, $S^{(n)}$ is constant then no further terms are generated. For instance, if $\vec{S}$ is constant then only the first order is required. We can view this as a gradient expansion which can be truncated if one is interested in the long wavelength physics of the system.   


\section{Anomalous Action} We turn now to calculating the anomalous contribution to the action. Following Fujikawa's method, we switch to Euclidean space and suppose that we have partially performed our transformation so that $\mathcal{S}_D\to\mathcal{S}_D(\tau)=\int d^4y \,\bar{\psi}(y) \slashed{D}(\tau)\psi(y)$ with $\tau\in [0,1]$ and 
\begin{equation}
\slashed{D}(\tau)=\gamma^\mu \left[\partial_\mu-i\tilde{A}_\mu(y;\tau)-i J(1-\tau)\tilde{S}_\mu(y;\tau)\right].
\end{equation}
Here we have introduced the partially rotated and rescaled field $\tilde{A}_\mu(y;\tau)$, (c.f. \eqref{tildeA}) which coincides with the original gauge field at $\tau=0$, $\tilde{A}_\mu(y;0)=A_\mu(x)$ and the final one at $\tau=1$, $\tilde{A}_\mu(y;1)=\tilde{A}_\mu(y)$. A similar definition is true for   $\tilde{S}_\mu(y;\tau)$. This partially rotated action coincides with the initial action, $\mathcal{S}_D$ and final action $\mathcal{S}'_D$ also at $\tau=0,1$ respectively. The anomalous contribution is found by considering an infinitesimal rotation such that $\mathcal{S}_D(\tau)\to \mathcal{S}_D(\tau+d\tau)$, calculating the Jacobian due to the transformation on the fields and then integrating this from $\tau=0$ to $\tau=1$. The result is~\cite{Fujikawa2004Book} 
\begin{equation}\label{Sanom}
 \mathcal{S}_\mathcal{A}= 2i\int_0^1 \!\!d\tau \!\!\int \!\!d^4x \Big\{\Omega(x;\tau)\text{Tr}[\mathbb{1}]+i\phi(x;\tau)\text{Tr}[\gamma_5]\Big\}
\end{equation}
 which is the sum of standard Weyl and chiral anomaly terms. Here the Tr[ ] denotes a trace over the Hilbert space as well as over spinor indices. The Hilbert space sum is naively divergent but can  be  regularized in the standard heat Kernel way,
\begin{equation}\label{Tr}
\text{Tr}[\mathcal{O}]=\lim_{M\to \infty}\int  \frac{d^4 k}{(2\pi)^4}e^{-ik_\mu x^\mu}\text{tr}\left[\mathcal{O}e^{-\slashed{D}^2(\tau)/M^2}\right]e^{ik_\mu x^\mu}
\end{equation}
with $\mathcal{O}=\mathbb{1},\gamma_5$ and tr[ ] denoting a trace over spinor indices only.   We note that in contrast to normal circumstances the generators of the chiral and Weyl transformations $\phi(x;\tau)$, $\Omega(x;\tau)$ themselves depend upon $\tau$. 

To calculate~\eqref{Tr} it is sufficient to expand the exponential up to at most fourth order as all other terms will be suppressed by the $M\to\infty$. After straightforward but tedious calculation we find  
\begin{eqnarray}\nonumber
\text{Tr}[\gamma_5]=iJ(1-\tau)\left[\frac{M^2}{4\pi^2}+\frac{[J(1-\tau)]^2\tilde{S}^2}{2\pi^2}-\frac{\partial_\mu\partial^\mu}{24\pi^2}\right]\partial_\alpha \tilde{S}^\alpha\\\label{gamma5}
+\frac{\epsilon^{\mu\nu\rho\sigma}}{8\pi^2}\left[\frac{[2J(1-\tau)]^2}{3}\partial_\mu \tilde{S}_\nu\partial_\rho \tilde{S}_\sigma
+e^2 \tilde{F}_{\mu\nu}\tilde{F}_{\rho\sigma}\right]~~~~
\end{eqnarray}
where  $\tilde{F}_{\mu\nu}=\partial_\mu\tilde{A}_\nu(y;\tau)-\partial_\nu \tilde{A}_\mu(y;\tau)$ and $\tilde{S}^2=\tilde{S}_\mu(y;\tau) \tilde{S}^\mu(y;\tau)$. The last term above is the standard  chiral anomaly term. A similar term also appears in the second line but is constructed purely from the spins. Amongst the remaining terms we note the divergent term $iJ(1-\tau)\frac{M^2}{4\pi^2}\partial_\alpha \tilde{S}^\alpha$ which we shall discuss further below. For the Weyl contribution we have
\begin{eqnarray}\nonumber
\text{Tr}[\mathbb{1}]&=&\frac{M^4}{4\pi^2}-\frac{J^2(1-\tau)^2}{24\pi^2}\bigg[12M^2 \tilde{S}^2+2\partial_\mu \tilde{S}_\nu \partial^\mu \tilde{S}^\nu-9\tilde{S}^4 \\ \label{one}
&& +4\tilde{S}_\mu \partial_\nu\partial^\nu \tilde{S}^\mu -\left( \partial_\mu \tilde{S}^\mu \right)^2\bigg] + \frac{e^2}{24\pi^2} \tilde{F}_{\mu\nu}\tilde{F}^{\mu\nu}
\end{eqnarray}
Again we see the presence of the usual Weyl anomaly contribution in  the first and last terms. The divergent term is typically discarded when considering the Weyl anomaly as it does not depend upon $\tilde{S}$ or $\tilde{A}$ but when calculating~\eqref{Sanom} it should be retained.



\section{Effective spin action} Combining \eqref{Sanom} with \eqref{gamma5} and \eqref{one} we have the exact anomalous action. To fully determine this requires us to perform the rather daunting seeming $\tau$ integral in \eqref{Sanom} which we do not attempt here. 
{{To get some understanding of what form this takes however we consider the case where the spin field takes the form
\begin{eqnarray}
\vec{S}(x)=\bar{\vec{S}}+\vec{\delta S}(x)
\end{eqnarray}
where $\bar{\vec{S}}$ is constant and $\vec{\delta  S}$ describe the fluctuations about this and then proceed by  computing the anomalous action using only the linearized solution~\eqref{M1},
\begin{eqnarray}\label{Apporximate}
U(x)=e^{i\gamma_5 J\left[\vec{x}\cdot \bar{\vec{S}}+\mathcal{G}*\delta\slashed{ S}\right]}.
\end{eqnarray}
The first term in the exponent is the standard chiral transformation used for Weyl semimetals which was discussed earlier and the  second arises due to the fluctuations. We now make the assumption that \eqref{Apporximate} provides a reasonable approximation to the transformation for the purpose of computing the low energy effective   anomalous action.}}
Using~\eqref{gamma5} we then find
\begin{align}\nonumber
\mathcal{S}_\mathcal{A}=-\!\!\!\int\! d^4x  \Bigg\{\!\frac{e^2J }{4\pi^2}\left[\bar{S}_\mu+\frac{1}{4}\partial_\mu \text{tr}\left(\mathcal{G}*\delta\slashed{ S}\right)\right]\!\epsilon^{\mu\nu\rho\sigma}A_\nu \partial_\rho A_\sigma\\\nonumber +\frac{J^3 }{18\pi^2}\left[\bar{S}_\mu+\frac{1}{4}\partial_\mu \text{tr}\left(\mathcal{G}*\delta\slashed{ S}\right)\right]\epsilon^{\mu\nu\rho\sigma}\delta S_\nu \partial_\rho\delta S_\sigma\quad\quad\\ \label{SpinAction}
 +\frac{J^2}{12\pi^2}\!\left[ \vec{\nabla}\!\cdot\vec{S}(x)\right]^2\ +\int d^4y\, S_i(x) V^{ij}(x-y)S_j(y)\!\Bigg\}\quad
\end{align}
where  $V_{ij}(z)=\mathcal{J}\partial_i\partial_j G(z)$. Adding this to $\mathcal{S}_\text{spin}$ we arrive at the approximate effective action for the spin system.  The first term here is the typical chiral anomaly term now modified to include the effect of the fluctuations, it represents a fermion mediated coupling of the spins to the gauge field.  The second has the same form as the first,  arising from a standard chiral anomaly term but built using spins.  Such a 3 spin term suggests a connection with the Wess-Zumino term occurring in the low energy action of fermions coupled to local moments~\cite{altland_simons_2010, GoswamiSi, Tsvelik, GoswamiSi2}. The third is a kinetic term for the spins generated from the coupling to the itinerant fermions.  Lastly, we have a long range RKKY interaction between the spins. The coupling constant  depends explicitly on the cutoff introduced earlier $\mathcal{J}=\frac{J^2 M^2}{2\pi^2}$~\footnote{
In \eqref{gamma5} a term $\sim \tilde{S}_\mu \tilde{S}^\mu\partial_\alpha \tilde{S}^\alpha$ is present. Since we are dealing with a spin system however $\vec{S}\cdot\vec{S}$ is a scalar of order one. This term contributes to $\mathcal{J}$ but it is negligible in comparison to $J^2M^2/2\pi^2$.}. 
The appearance of this divergence is natural in models such as ours and is akin to the well known divergence of the vacuum polarization in QED which is governed by the same set of diagrams. In a condensed matter context, deviations from a linear dispersion will cure this divergence giving a finite but non-universal coupling constant. From this we can determine the leading order renormalization group (RG) flow of this RKKY coupling $\d{\mathcal{J}}{ l}=2\mathcal{J}$ with $l=\log{M}$ indicating it is relevant in an RG sense. 

{{If we were to include terms beyond the linear approximation in our transformation then this would result in 4 spin terms as well as terms involving higher derivative terms, which are typically dropped when computing an effective action. For these reasons we content ourselves with the linearized approximation but note that the presence of the Weyl transformation at higher orders provides a means to determine the RG flow of the terms present in \eqref{SpinAction}. }}


\section{Transport} 
{{We turn our attention now to calculating the anomalous transport in the system which we will be able to do without resorting to approximations as done in the previous section.}}  In principle,  this requires evaluating the integral \eqref{Sanom} fortunately  however, this turns out to be not necessary. To see this we note that the anomalous current is found by varying $\mathcal{S}_\mathcal{A}$ with respect to $A_\mu(x)=\tilde{A}_\mu(x,\tau=0)$. Thus 
\begin{eqnarray}\nonumber
\left<j^\mu(x)\right>&=&\pd{\mathcal{S}_\mathcal{A}}{\tilde{A}_\mu(x,0)}=-2\phi^{(1)}(x)\pd{\text{Tr}[\gamma_5]|_{\tau=0}}{A_\mu(x,0)}
\end{eqnarray}
where the second equality follows from the fact that the variation is carried out at $\tau =0$ along with  $\phi(x,0)=\phi^{(1)}(x)$ and $\Omega(x,0)=\Omega^{(1)}(x)=0$. From this we find the density response to be $\rho(x)=\frac{e^2}{2\pi^2}\vect{\nabla}\phi^{(1)}(x)\cdot \vec{B}$ or in Fourier space,
\begin{align}
\rho(\vec{q},\nu)=\frac{e^2J}{2\pi^2}\int_{\vec{k}\omega}\frac{k_i k_j }{|k|^2-\omega^2}\left<S^j(\vec{k},\omega)\right>_SB^i(\vec{q}-\vec{k},\nu-\omega)
\end{align}
where we have used the shorthand $\int_{\vec{k}\omega}=\int d^3k \,d\omega/(2\pi)^4$ and $B^i(\vec{k},\omega)$ is the applied magnetic field in Fourier space. The expectation value on the right is taken with respect to the effective spin action~\eqref{SpinAction} or alternatively could represent some imposed, mean field spin configuration. This generalizes the result for a Weyl semimetal to the case of non constant $\vec{S}(x)$.  It describes the response of the system to a density perturbation in the presence of an arbitrary magnetic field.   

Similarly the current is 
\begin{eqnarray}
\vec{j}(x)=\frac{e^2}{2\pi^2}\vect{\nabla}\phi^{(1)}\!\times \vec{E}-\frac{e^2}{2\pi^2}\partial_t\phi^{(1)}\vec{B}
\end{eqnarray} or in Fourier space,
\begin{eqnarray}\nonumber
j^l(\vec{q},\nu)=\frac{e^2J}{2\pi^2}\int_{\vec{k}\omega}\frac{ k_i\left<S^j(\vec{k},\omega)\right>_S}{|k|^2-\omega^2} \left[ \epsilon^{ljs}k_j E_s(\vec{q}-\vec{k},\nu-\omega)\right.\\
\left.+\omega B^l(\vec{q}-\vec{k},\nu-\omega)\right]~~~~~~~~~
\end{eqnarray}
In the first line we can recognize the generalization of the standard Hall current expression to the case of non constant $\vec{S}(x)$. In addition to this we note the presence of the magnetic field which gives rise to a chiral magnetic effect. This is in contrast to the simple Weyl case discussed above wherein the CME requires that $S_0\neq 0$ which can be the case in the absence of inversion symmetry. 
{{This in turn results in a time dependent chiral rotation $\sim e^{iJ\gamma_5 S_0t}$ and a corresponding term in the anomalous action. In the current circumstances, although $S_0=0$ and the symmetry is not broken a CME is still generated via the time dependent nature of the transformation.}}



\section{Discussion \& Conclusion}  
{{ In this paper we have presented an alternative approach to interacting semimetals based on the technique of functional bosonization from $1+1$ dimensions, generalized to $3+1$ dimensions. We have focused here on the case of a Kondo-semimetal wherein the semimetal is coupled to an array of spins, although our method can be  applied to strained semimetals also. Our method relies on the existence of a non-Abelian transformation of the fermions which decouples them from the spin system. This transformation is anomalous, due to the presence of the chiral and Weyl anomalies, and by calculating its non-trivial Jacobian the low energy effective action for the spin system can be determined in addition to the anomalous transport. 

This approach can also be used for the evaluation of correlation functions. For instance the fermionic Green's function is given by
\begin{eqnarray}\nonumber
 i\left<\psi_\alpha(x)\bar{\psi}_\beta(0)\right>&=&\left<U_\alpha^{\alpha'}(x)\bar{U}_\beta^{\beta'}(0)\right>_{S}\mathcal{G}_{\alpha'\beta'}(x)
\end{eqnarray}
where once again $\left<\right>_S$ denotes the expectation value with respect to the spin system only  and we have restored spinor indices. The factorization of the correlation functions into a free fermionic part, $\mathcal{G}(x)$, and a bosonic part is a hallmark of the bosonization method and in (1+1)-d  provides a simple route to finding non-Fermi liquid  behaviour~\cite{Naon, LeeChen, GiamarchiBook, GogolinNerseyanTsvelikBook}.  Upon adopting the linear approximation $U(x)\approx e^{i\gamma_5 J\mathcal{G}*\slashed{S}(x)}$ this expression simplifies and the exponential form of the spin factor can facilitate  evaluation of the expectation value and, potentially non-Fermi liquid  correlations.   }}


\acknowledgements  This work was supported by the U.S. Department of Energy, Office of Science, Basic Energy Sciences under Award No. DE-SC0001911 and the Simons Foundation.


\bibliographystyle{apsrev4-1}
\bibliography{Main3.bib}

\end{document}